\documentclass[aps,prc,a4paper,nofootinbib,preprintnumbers,superscriptaddress,twocolumn,amssymb,amsmath]{revtex4-1}

\usepackage{mathtools}
\usepackage{slashed}
\usepackage{epsfig}
\usepackage{bbold}
\usepackage{ulem}
\usepackage{bm}
\usepackage{commath}

\usepackage{xcolor}
\usepackage{ulem}
\definecolor{purple}{rgb}{0.8,0,0.6}
\definecolor{darkslategray}{rgb}{0.18,0.31,0.31}

\renewcommand\sout{\bgroup \color{red} \ULdepth=-.5ex \ULset}

\newcommand{\beqn}{\begin{eqnarray}}
\newcommand{\eeqn}{\end{eqnarray}}
\newcommand{\beqs}{\begin{subequations}}
\newcommand{\eeqs}{\end{subequations}\\[-2mm]\noindent}
\newcommand{\eq}[1]{(\ref{#1})}
\newcommand{\bs}{\boldsymbol}
\newcommand{\bx}{{\boldsymbol x}}
\newcommand{\bp}{{\boldsymbol p}}
\newcommand{\bv}{{\boldsymbol v}}
\newcommand{\bl}{{\bs \ell}_\bp}
\newcommand{\cA}{{\mathcal A}}

\begin{document}

\title{Conformal anomaly and helicity effects in kinetic theory via scale-dependent coupling}

\author{M. N. Chernodub}
\affiliation{Institut Denis Poisson UMR 7013, Universit\'e de Tours, 37200 France}
\affiliation{Pacific Quantum Center, Far Eastern Federal University, Sukhanova 8, Vladivostok, 690950, Russia}

\author{Eda Kilin\c{c}arslan}
\affiliation{Physics Engineering Department, Faculty of Science and Letters, Istanbul Technical University, Maslak--Istanbul, 34469, Turkey}

\date{\today}

\begin{abstract}
We modify a kinetic theory of massless fermions to incorporate the effects of the conformal anomaly. Working in a collisionless regime, we emulate the conformal anomaly via a momentum-dependent electric coupling. In this prescription, the conformal anomaly leads to a hedgehog-like structure in the momentum space similar to the Berry phase associated with the axial anomaly. The interplay between the axial and conformal anomalies generates the axial current, proportional to the helicity flow of the electromagnetic background. The corresponding conductivity is determined by the running of the electric coupling between the tip of the Dirac cone and the Fermi surface.
\end{abstract}

\maketitle

\section{Introduction}

Chiral fermions appear in different physical environments ranging from quark-gluon plasma created in heavy-ion collisions~\cite{Kharzeev:2008lw, Kharzeev:2008fw} to the electronic excitations in Weyl semimetals in the solid-state physics~\cite{Turner:2011vs,Burkov:2011b,Yang:2011lr}. Major properties of these systems carry an imprint of the chiral invariance respected by the classical theory of the massless spin-half particles. The invariance implies that the left-handed and right-handed fermionic currents are separately conserved.

Even though the massless Dirac equation is invariant under chiral transformations, the chiral symmetry of the classical theory is broken due to quantum mechanical effects in the presence of the external electromagnetic fields. This phenomenon, known as the quantum axial anomaly, affects the transport properties of the chiral medium leading to the experimentally accessible signatures such as the chiral magnetic effect (CME)~\cite{Kharzeev:2008lw, Kharzeev:2008fw, Kharzeev:2007z} and the chiral separation effect~\cite{Metlitski:2005z, Jensen:2013kr} that generate vector (electric) and axial currents along the direction of the background magnetic field. The mixed axial-gravitational anomaly~\cite{Landsteiner:2011cp} leads to an anomalous magneto-thermoelectric transport in the magnetic-field background parallel to a temperature gradient~\cite{Chernodub:2013kya,Gooth:2017}. 

The basic physics of the chiral fermionic systems is described by a massless QED which incorporates the breaking of the Lorentz invariance in real materials. Due to the absence of massive parameters for the ungapped quasiparticle modes, the corresponding infrared effective models possess a scale (conformal) invariance at the classical level. As a result, the classical processes look identically the same under a suitable rescaling of coordinates, energies and momenta according to their canonical dimensions. 

Similarly to the axial symmetry, the conformal symmetry is broken at the quantum level by the conformal anomaly. At the level of the transport properties, the conformal anomaly generates the scale magnetic effect~\cite{Chernodub:2016lbo} which was suggested to generate an anomalous thermoelectric transport~\cite{Chernodub:2017jcp,Arjona:2019lxz}. The transport effects induced by the conformal anomaly may also play a role in the heavy-ion collisions~\cite{Kawaguchi:2020kce}.

Dynamic evolution of a large system of particles may be described with the help of an appropriate kinetic theory. In Ref.~\cite{CME:SY}, the kinetic theory for chiral fermions exhibiting the CME was constructed using the path integral formalism. The chiral kinetic equation incorporating the axial anomaly into the classical kinetic theory of Weyl fermions was proposed in Refs.~\cite{SonY:2012,Son:2012zy} and elaborated further in Refs.~\cite{Chen:2013s.pww, Manuel-Torres:2014-2,chen2014:ssyyLI,Manuel-Torres:2014, chen2015:ssyI,Eda:2017dy,Gorbar:2017mss,Abbasi:2018zoc} in different approaches. The relativistic chiral kinetic theory has also been developed in the framework of the Wigner functions \cite{Hidaka:2017py,Hidaka:2017py-2,Huang:2018slz, Carignano:2018ct-r,Eda:2018d,Nora:2019sswk} as well as within the scope of the worldline formalism~\cite{Mueller:2018v}. The chiral kinetic theory in curved spacetime has been considered in Ref.~\cite{Liu:2018xip} while the effects of a finite viscosity on anomalous transport of the chiral fluids has been addressed in Ref.~\cite{Hidaka:2018l}. 

The anomalous transport phenomena considered so far in the scope of the kinetic theory were related to the axial anomaly or the mixed axial-gravitational anomaly. In our paper, we make an attempt to incorporate the conformal anomaly at the level of the classical kinetic theory of massless fermions. For simplicity, we consider the collisionless limit and focus on the scale symmetry breaking associated with the running of the electric coupling. 

The plan of this paper is as follows. In Section~\ref{sec:kinetic} we first review the single quasiparticle theory in a collisionless limit. Then we discuss how the conformal anomaly may appear in the theory where the collisions are absent. We also highlight the subtleties of delegating of the anomalous scaling properties to the momentum-dependent coupling. We end this section with the derivation of the kinetic equation that includes both axial and conformal anomalies. In the second part of the paper, in Section~\ref{sec:transport}, we explore the new transport phenomena that appear due to the scale violation. Our last section is devoted to the conclusions where we also discuss the limitations of our approach and realization of the proposed transport effects in physical systems.

\section{Kinetic theory of chiral fermions}
\label{sec:kinetic}

\subsection{Single quasiparticle theory}

We consider a system of relativistic chiral fermions described, in the absence of background fields, by the following equation:
\beqn
({\bs \sigma}\cdot \bp) u_\bp = \pm p\, u_{\bp},
\label{eq:Weyl}
\eeqn
where the upper and lower signs correspond, respectively, to the right- and left-handed two-component Weyl spinors~$u_{\bp}$ carrying the momentum ${\bs p}$ (with $p \equiv |{\bs p}|$). The fermions propagate with the Fermi velocity which is set to unity $v_F = 1$ for simplicity.

It is useful to associate to the wave function of the Weyl fermion~\eq{eq:Weyl} a fictitious vector potential, called the Berry connection, ${\bs \cA}_\bp$, and the corresponding fictitious magnetic field (the Berry curvature) ${\bs \Omega}_{\bp}$, both defined in the momentum space~\cite{Berry:1984jv}:
\beqn
{\bs \cA}_\bp = i {u_\bp}^{\hskip -2mm \dagger \hskip 0.5mm} {\bs \nabla}_\bp u_\bp,
\qquad
{\bs \Omega}_{\bp} \equiv {\bs \nabla}_\bp \times {\bs \cA}_{\bp} = s \frac{\bp}{p^3}.
\label{eq:A:Omega}
\eeqn
Here $s$ is the spin of the particle. Since the right- and left-handed fermions with half-spin particle possess the Berry curvatures of opposite signs, it is convenient to allow for the spin $s$ to carry the sign of the appropriate helicity, $s= + 1/2$ and $s = - 1/2$, respectively. In our work, we will concentrate mostly on the right-handed fermions and include both helicities at the very end. 

The second formula in Eq.~\eq{eq:A:Omega} is also valid for a scalar field with a zero spin $s=0$. As the scalar field possess a trivial Berry curvature, ${\bs \Omega}_{\bp} = 0$, we will use the scalar particle to discriminate the effects of axial and conformal anomalies. 

The fictitious magnetic field~\eq{eq:A:Omega} has a form of a hedgehog that resembles a ‘‘magnetic monopole’’ in the momentum space with the center at the origin, ${\bs p} = 0$. In a finite-density fermionic system with $\mu \neq 0$, the Berry curvature carries a quantized ``magnetic'' flux through the Fermi surface which corresponds to the quantization of the anomalous vertex that appears due to the axial anomaly.

The action of a single electrically charged relativistic particle in the presence of the background electromagnetic field $A^\mu = A^\mu({\bs x})$ has the following form~\cite{ref:Xiao,Duval:2005vn}:
\beqn
S {=} {\int} d t \left[ p^i {\dot x}^i {+} e A^i({\bs x}) {\dot x}^i {-} \cA^i({\bs p}) {\dot p}^i {-} \epsilon_{\bs p}({\bs x}) {-} e A^0({\bs x}) \right],\quad\
\label{eq:S:1}
\eeqn
where $e$ is the electric coupling of the quasiparticle and the dot over a symbol indicates a partial derivative over time (${\dot {\bs x}} = \partial {\bs x} / \partial t$, etc). 

The form of the action~\eq{eq:S:1} implies that the motion of chiral fermions is affected both by the electromagnetic field $A^\mu = A^\mu(x)$ in the coordinate space and by the Berry curvature $\cA^i = \cA^i(p)$ in the momentum space. The inclusion of the Berry curvature incorporates the effect of the axial triangle anomaly and, consequently, leads to the anomalous transport phenomenon such as the CME~\cite{CME:SY,Son:2012zy,SonY:2012}.

The action~\eq{eq:S:1} can be cast into the suggestive form:
\beqn
S= \int d t \left[ - \omega_a(\xi) {\dot \xi}^a - H(\xi) \right],
\label{eq:S:2}
\eeqn
where the quantity $H(\xi) = \epsilon_{\bs p}({\bs x}) + e A^0({\bs x})$ can readily be associated with a Hamiltonian. In Eq.~\eq{eq:S:2}, the space coordinates ($x^i$, $i=1,2,3$) and the momentum coordinates ($p^i$, $i=1,2,3$) are combined into the single vector with the six components: $\xi^a = (x^1, \dots, p^3)$ with $a = 1,\dots 6$. The action~\eq{eq:S:2} gives rise to the following equations of motion:
\beqn
\omega_{ab} {\dot \xi}^b = - \partial_a H, \qquad
\label{eq:EOM:1}
\eeqn
where $\omega_{ab} = \partial_a \omega_b - \partial_b \omega_a$ and $\partial_a \equiv \partial / \partial \xi^a$. Associating the inverse $\omega^{ab} \equiv \left( \omega^{-1} \right)^{ab}$ with the Poisson brackets
\beqn
\left\{ \xi^a, \xi^b \right\} = \omega^{ab},
\label{eq:Poisson}
\eeqn
we rewrite the equations of motion for the quasiparticle in the canonical form:
\beqn
{\dot \xi}^a = - \omega^{ab} \partial_b H = \left\{ H, \xi^a \right\} = - \left\{ \xi^a,\xi^b \right\} \frac{\partial H}{\partial \xi^b}.
\label{eq:EOM:2}
\eeqn
These equations describe the motion of the fermionic electrically charged particle in a background of the electromagnetic field. The aim of our paper is to figure out how the running of the electric coupling affects the dynamics of the fermionic ensembles. In the next section, we discuss how the conformal anomaly associated with the renormalization of the electric coupling can be incorporated in the above derivation.

\subsection{Mimicking the conformal anomaly}

In a conformally invariant field theory, processes at different energy and length scales are described by identical classical equations of motion. The conformal anomaly appears naturally at the quantum level because, in a general theory, quantum corrections affect differently the physical phenomena that develop at different scales. The presence of the conformal anomaly reveals itself in the fact of the ``running'' of a coupling which implies the dependence of the coupling with respect to the energy scale $\varepsilon$. In simple terms, in an interacting quantum field theory, the coupling constants are not, strictly speaking, constants. 

Contrary to the classical theory, particles that scatter at each other at different energies interact with different couplings. In particular, quantum fluctuations lead to the running of the electric coupling in quantum electrodynamics: at large distances, the electric field of a test electron gets screened by quantum fluctuations that create virtual pairs of electrons and positrons. Therefore, a pair of electrically charged particles that interact at higher energy, would approach each other closer and interact with a higher effective electric coupling as compared to a less energetic pair of particles. 

The running of couplings appear, for example, in a QED-like theory of (relativistic) electrically charged fermions that describes topological insulators and semimetals. In this environment, the Lorentz invariance is broken by the presence of the material. In particular, the velocity of quasiparticles $v$ does not coincide with the velocity of photons $c$ in the medium. As a result, in addition to the running electric coupling\footnote{To keep our notations concise, in this article the symbol $e$ denotes also the electric coupling of the quasiparticles, $e \equiv e^*$.} $e$, the velocities of photons and quasiparticles become dependent on the energy scale, contrary to the renormalization properties of the Lorentz-invariant QED. In the isotropic crystals, these velocities approach each other in the infrared fixed point and the Lorentz symmetry gets partially restored~\cite{Isobe:2012vh} (see also~\cite{Roy:2015zna}). In anisotropic Weyl and Dirac semimetals, in which the relativistic energy-dispersion cones are tilted, the anisotropy surprisingly persists even at the infrared fixed point~\cite{Vozmediano:2018pf}. 

A slightly different picture emerges in two spatial dimensions. In graphene, the electric coupling is not renormalized at all while the Fermi velocity runs with the interaction scale~\cite{Gonzalez:1993uz,Vozmediano:2010fz}. The running of the velocity was observed experimentally~\cite{ref:Nature:renormalization} with the running scale fixed by the Fermi level of electrons in suspended graphene: a change of the Fermi level $\mu$ results in a change of the Fermi velocity, $v = v (\mu)$, according to the renormalization group~\cite{Gonzalez:1993uz}. The velocity renormalization has also been verified in numerical simulations~\cite{Stauber:2017fuj}.

In the fundamental QED, the products of the electric coupling $e$ and the strengths of magnetic $B$ or electric $E$ background fields are the renormalization-invariant quantities~\cite{Schwinger:1951nm,Dunne:2004nc}: $e_{\mathrm{ren}} B_{\mathrm{ren}} = e_0 B_0$ and $e_{\mathrm{ren}} E_{\mathrm{ren}} = e_0 E_0$. Therefore, in a consistent (fundamental) field theory such as QED, the combinations $e {\bs A}$ and $e A^0$ in the quasiparticle action~\eq{eq:S:1} should be considered as renormalization-group invariants. This property means that the conformal anomaly cannot enter the collisionless kinetic theory of a fundamental model described by the particle action similar to Eq.~\eq{eq:S:1}. 

In our paper, we examine the case when the renormalization of the electric coupling $e$ of a particular type of quasiparticles does not match the renormalization of the electromagnetic field. For example, the physical spectrum may contain other electromagnetically active excitations, which affect the background magnetic field and renormalize the effective coupling of the quasiparticle differently. 

\begin{figure}[htb]
\center{\includegraphics[width=0.6\linewidth]{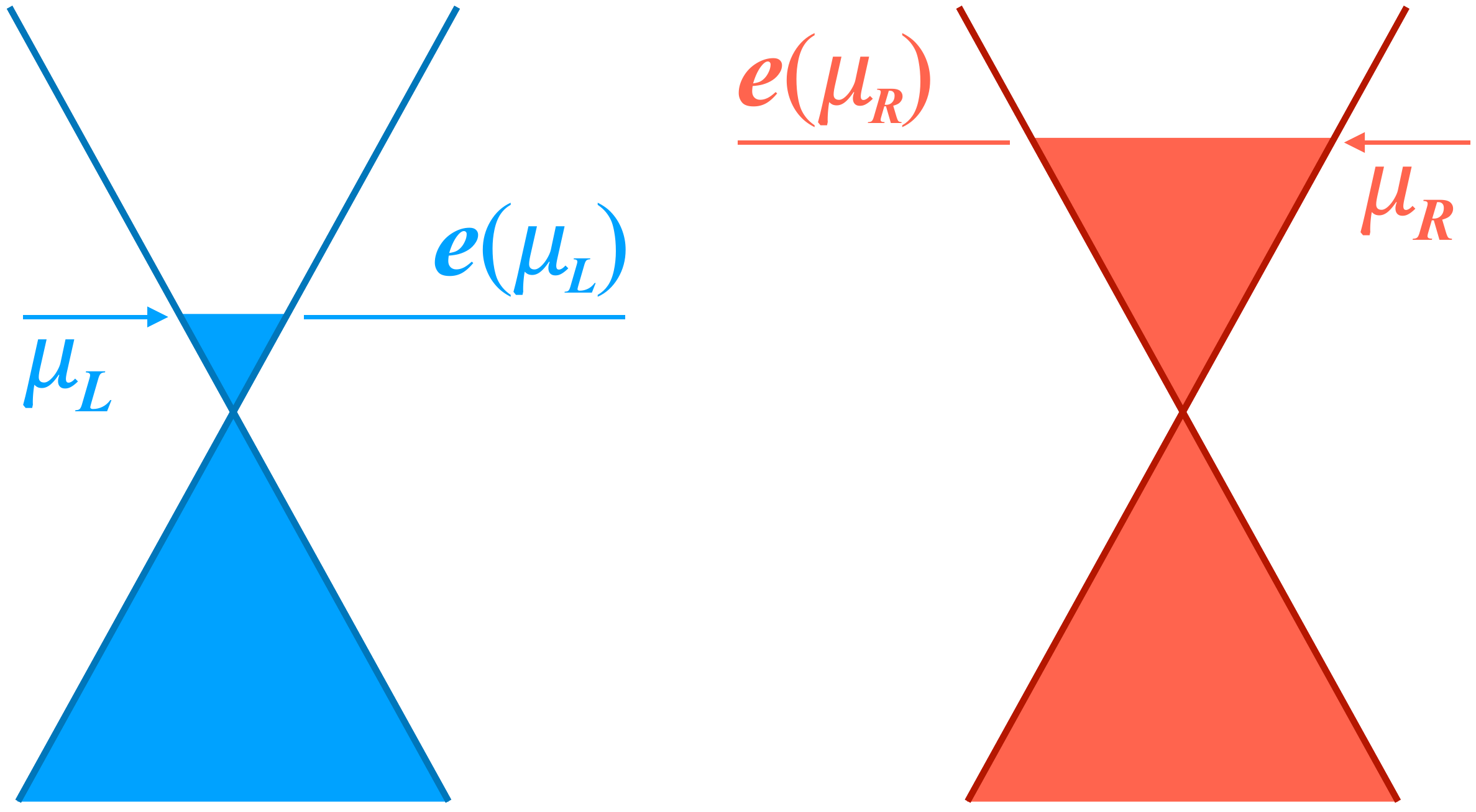}}
\caption{The running of the electric coupling at the Fermi surfaces of the left-handed and right-handed fermions.}
\label{fig:illustration}
\end{figure}

In such theories, the conformal anomaly could also interfere with the axial anomaly. In the magnetic field background, the axial anomaly leads to the chiral magnetic effect~\cite{Kharzeev:2008fw}: the generation of vector (electric) current in a presence of the axial (chiral) imbalance. In a chirally imbalanced system, the (quasi)particles with right- and left-handed chiralities possess different Fermi levels, $\mu_R$ and $\mu_L$, respectively, with $\mu_R \neq \mu_L$. Due the running of the electric coupling $e = e(\mu)$,  electric couplings $e$ of the right- and left-handed fermions take different values, $e_R = e(\mu_R)$ and $e_L = e(\mu_L)$, with $e_L \neq e_R$ if $\mu_R \neq \mu_L$ (see Fig.~\ref{fig:illustration} for an illustration). Since the magnetic field $B$ takes the same value for particles of both chiralities, the effect of the magnetic field on different chiralities, expressed via the products $e_{R} B(\mu_R)$ and $e_L B(\mu_L)$, could be different, $e_{R} B(\mu_R) \neq e_L B(\mu_L)$, thus resulting in renormalization of the CME conductivities in such theories. The origin of this effect is similar to the renormalization of the Fermi velocity in graphene which was mentioned earlier. 

In other words, we probe the effective infrared theory of the excitations for which the combinations $eB$ and $eE$ are no more renormalization-group invariants for certain relativistic degrees of freedom, having in mind, as an example, right-handed and left-handed fermionic quasiparticles in semimetals. One could speculate that this situation can be realized, for example, in the cases when the system possesses, in addition to the fermionic quasiparticle, other excitations, that renormalize the background electromagnetic fields. 

For the sake of simplicity, in our paper we consider an isotropic material. We do not explicitly include the permittivity and the permeability of the material, setting both of them to unity. Moreover, to get our presentation as clear as possible, we assume that the background fields and the quasiparticle velocity are not renormalized and the effect of the running is featured in the electromagnetic coupling only. 

The electric coupling $e$ is a function of momentum $p$ transferred in the interaction event. The running of the electric coupling is controlled by the corresponding beta function
\beqn
\beta_e = \frac{d e(\varepsilon)}{d \ln \varepsilon} \equiv \frac{d e(p)}{d \ln p}.
\label{eq:beta:g}
\eeqn
One can also define the running of electric coupling in terms of the related beta function 
\beqn
\beta_{\alpha} = \frac{d \alpha(p)}{d \ln p} \equiv \frac{e}{2 \pi} \beta_e,
\label{eq:beta:alpha}
\eeqn
for the QED-like fine structure constant\footnote{We remind that we set the permittivity of the material to unity.} $\alpha = e^2/(4\pi)$. Below we will use both functions~\eq{eq:beta:g} and \eq{eq:beta:alpha}.

In our paper, we work in the collisionless limit of the kinetic theory where the collision integral is zero and the interactions between particles are ignored. Therefore, strictly speaking, the collisionless theory contains no room for inter-particle interactions and has, naively, no chance for the conformal anomaly to play any role. However, the kinetic theory is an effective theory which gives us a freedom to incorporate quantum anomaly effects via the appropriate modification of the action. For example, the axial anomaly is taken into account by the new term with the Berry connection in the action~\eq{eq:S:1} which is not present in the original classical theory. 

We will use the same opportunity to mimic the effects of the scale anomaly: we associate the momentum in Eqs.~\eq{eq:beta:g} and \eq{eq:beta:alpha} with the momentum of the particle with respect to the background (thermal) frame, where the equilibrium thermal bath is defined. In this Lorentz frame, the chemical potential has only a time-like component. We consider the theory in the background of a uniform electromagnetic field $({\bs E},{\bs B})$ defined in the same frame. In order to capture the effects of the conformal anomaly for the dynamics of a single charged quasiparticle, we assume that the electric coupling is a function of the absolute value of the quasiparticle momentum~$p$ with respect to the background frame~$e = e(p)$. As a consistency check, we will make sure at the end of our calculations that our results are given by Lorentz-covariant expressions. 

\subsection{Kinetic equation with axial and conformal anomalies}

In the vacuum of massless QED, the conformal anomaly associated with the electric coupling reveals itself in the form of anomalous transport phenomena, the scale electromagnetic effects, that are realized in the background of gravitational and electromagnetic fields~\cite{Chernodub:2016lbo}. While these effects were predicted to exist in the vacuum, we will explore whether similar effects are experienced by relativistic matter. In the presence of the running electric coupling, the action~\eq{eq:S:1} takes the following form:
\beqn
S = \int d t  &&\left[  p_i {\dot x}^i  + e(p) A^i({\bs x}) {\dot x}^i  \right.\nonumber \\
& & \left. - \cA^i({\bs p}) {\dot p}^i - \epsilon_{\bs p}({\bs x}) - e(p) A^0({\bs x}) \right].\quad\
\label{eq:S:3-2}
\eeqn

The $6 \times 6$ matrix $\omega_{ab}$ in the equation of motion~\eq{eq:EOM:1} has the $3 \times 3$ block form:
\beqn
\omega_{ab}  =
\left(
\begin{array}{c|c}
\omega_{\bs F} & \omega_{\beta} \\
\hline
-\omega_{\beta} & \omega_{\bm C}
\end{array}
\right),
\label{eq:omega:ab}
\eeqn
where
\beqn
\left( \omega_{\bs F} \right)_{ij}  & = & - e F_{ij}
\nonumber \\
\left( \omega_\beta  \right)_{ij} & = & -  \delta_{ij} - \beta_e p_i A_j /p^2, 
\label{eq:running:omega} \\
\left( \omega_{\bs C} \right)_{ij} & = & \epsilon_{ijk} \Omega_{\bm p_k}. \nonumber 
\eeqn
The electromagnetic field background enters via the classical field-strength tensor $F_{ij} = \partial_i A_j - \partial_j A_i$. The running of the electric coupling~\eq{eq:beta:g} appears in block-off-diagonal terms $\omega_\beta$ of the matrix~\eq{eq:omega:ab} given in Eq.~\eq{eq:running:omega}.

A straightforward calculation of the Poisson brackets~\eq{eq:Poisson} gives:
\beqn
\left\{ x^i, x^j \right\}  & = & \frac{1}{\sqrt{w}} \epsilon^{ijk} \Bigl[ \Omega_k + (\bm \Omega_{\bm p} \cdot \bl) A_k \Bigr], 
\label{eq:Poisson:conf}\\
\left\{ p^i, x^j \right\}  & = & \frac{1}{\sqrt{w}} \biggl[ \delta^{ij}(1+\bm A \cdot \bl) -  A^i \ell^j  +  \Omega^i (e(p) B^j) \biggr], \qquad 
\nonumber \\
\left\{ p^i, p^j \right\}  & = & - \frac{1}{\sqrt{w}} \epsilon^{ijk} \Bigl[ e(p) B_k +  \ell_k {\bs A} \cdot {\bs B} \Bigr]\,,
\nonumber
\eeqn
where $w = \det \omega_{ab}$ and ${\bs B} = {\bs \nabla} \times {\bs A}$ is the magnetic field.

Equations~\eq{eq:omega:ab}--\eq{eq:Poisson:conf} indicate that the running electric coupling enters the kinetic theory via the single combination of the beta function and the momentum:
\beqn
\bl  = \beta_e(p) \frac{\bp}{p^2}.
\label{eq:l}
\eeqn
Surprisingly, the ``conformal vector''~\eq{eq:l} has a similar monopole hedgehog structure as the Berry curvature~\eq{eq:A:Omega}, $\bl \sim {\bs \Omega}_{\bs p}$ which has been introduced to the kinetic theory to mimic the axial, not conformal, anomaly~\cite{CME:SY,Son:2012zy,SonY:2012}. 

Contrary to the ``axial monopole'', the total flux of the conformal hedgehog (more precisely, $\bl/p$) is not quantized since the integral of the conformal vector~\eq{eq:l} over the Fermi surface gives us a quantity proportional to the beta function evaluated at the Fermi surface, $\beta_e = \beta_e( p = \mu)$. Therefore the ``conformal flux'' depends on the radius $\mu$ of the surface.

The conformal vector~$\bl$ is sensitive to the beta function of the theory and insensitive on (the sign of) the particle's spin. Therefore, the effects of the conformal anomaly may appear for a spinless bosonic particle ($s = 0$) contrary to the effects of the axial anomaly. The Berry curvature~${\bs \Omega}_{\bs p}$, instead, is insensitive to the running coupling while being dependent on the handedness of the particle. 

In a complete analogy with the effect of a Berry curvature~\cite{ref:Xiao,Duval:2005vn}, the conformal anomaly modifies the measure in the phase space:
\begin{eqnarray}
{\mathrm{d}} \Gamma = \sqrt{w} {\mathrm d} \xi & \equiv & 
\biggl[1 + {\bs \ell}_\bp \cdot \bm A + e(p) {\bs \Omega}_{\bp}  \cdot {\bm B} \nonumber\\
&&  + e(p) ({\bs \ell}_\bp \cdot {\bs \Omega}_{\bp}) \ \bm A  \cdot {\bm B} 
\biggr] \frac{d \bx d \bp}{(2\pi)^3}.\quad \label{eq:Gamma}
\end{eqnarray}
Both the Poisson brackets~\eq{eq:Poisson:conf} and the measure~\eq{eq:Gamma} contains gauge-variant contributions coming from the gauge field alone. These contributions are always associated with the conformal vector $\bl$. We will see below that the final expressions for physical currents will have well-defined physical meaning despite the gauge dependence of the intermediate expressions. The appearance of some of these quantities may be traced back to the equations above, \eq{eq:Poisson:conf} and \eq{eq:Gamma}, where we find, also unexpectedly, the density of the (magnetic) helicity given by the Chern-Simons term:
\beqn
K^0 = {\bs A} \cdot {\bs B} \equiv {\bs A} \wedge {\bs F} \equiv \frac{1}{2} \epsilon^{ijk} A_i F_{jk},
\label{eq:AB}
\eeqn
The topological terms of the kind~\eq{eq:AB} are usually encountered in systems with axial, but not conformal, anomalies.

The integral over the density of the helicity~\eq{eq:AB} gives us the total magnetic helicity:
\beqn
H = \int d^3 x \, {\bs A} \cdot {\bs B}.
\label{eq:helicity}
\eeqn
This quantity is an important dynamical invariant of ideal magnetohydrodynamics. It plays an essential role in physical applications ranging from laboratory plasmas to astrophysical systems~\cite{Brown:1999}. 

Physically, the total helicity number~\eq{eq:helicity}, understood as an expectation value over a photon state, counts the difference between the numbers of the left- and right-handed photons in the electromagnetic field. It may also be interpreted as a topological linking number of the magnetic field lines~ \cite{Moffatt:1969}. This interpretation becomes explicitly transparent in the Coulomb gauge ${\bs \nabla} \cdot {\bs A} = 0$, where the helicity~\eq{eq:helicity} was shown to be related to flux-weighted Gauss linking integral calculated at (and averaged over) all pairs of magnetic field lines~\cite{Renzo:2011}. 

In the absence of the conformal anomaly ($\beta_e = 0$), the Poisson brackets~\eq{eq:Poisson:conf} and the phase element~\eq{eq:Gamma} reduce to the familiar expressions, respectively:  
\beqn
\left\{ x^i, x^j \right\} & = & \epsilon^{ijk}  \frac{\Omega_k}{\sqrt{w}}, \nonumber\\
\qquad \left\{ p^i, p^j \right\} & = & - e(p) \epsilon^{ijk} \frac{B_k}{\sqrt{w}}, \qquad 
\label{eq:Poisson:standart} \\
\left\{ p^i, x^j \right\} & = & \frac{\delta^{ij} + e(p) B^j \Omega^i }{\sqrt{w}}, \nonumber\\
\mathrm{d} \Gamma\equiv{\mathrm d} \xi & = & (1+e(p) \bm B \cdot {\bs \Omega}_{\bp}) \frac{d \bx d \bp}{(2\pi)^3}.
\nonumber
\eeqn

The requirement of the conservation of the occupation number, $d n/ d t = 0 $, leads us to the (collisionless) kinetic equation:
\beqn
{\dot n}_\bp + \omega^{ab} \partial_a H \partial_b n_\bp = 0\,,
\eeqn
which can be written in the following explicit form:
\begin{eqnarray} \begin{aligned}
 &{\dot n}_\bp +  \frac{1}{\sqrt{w}}  \Big\{ \Bigl(   e {\widetilde {\bs E}} +  {\bs \ell}_\bp \times [e {\widetilde {\bs E}} \times {\bs A}] +  {\tilde \bv}_\bp \times e {\bs B}  \\
 & +e {\bs B}  \cdot \bm A \ ({\tilde \bv}_\bp \times {\bs \ell}_\bp) +   (e {\bs B} \cdot e {\widetilde {\bs E}}) \ {\bs \Omega}_{\bp} \Bigr) \cdot \frac{\partial n_\bp}{\partial \bp}  \\
 & \Bigr(  {\tilde \bv}_\bp +  {\bs A} \times [ {\tilde \bv}_\bp \times \bl ] +  e {\widetilde {\bs E}} \times {\bs \Omega}_{\bp} \\
 & + e ({\bs \Omega}_{\bp} \cdot {\bs \ell}_\bp) {\widetilde {\bs E}} \times \bm A + ({\bs \Omega}_{\bp} \cdot {\tilde \bv}_\bp) e {\bs B} )  \Bigr) \cdot \frac{\partial n_\bp}{\partial \bx} \Big\}  = 0,
 \label{eom}
\end{aligned}\end{eqnarray}
where 
\beqn
\sqrt{w}=1 +  {\bs \ell}_\bp \cdot \bm A +  e {\bs B} \cdot {\bs \Omega}_{\bp} \nonumber + {\bs \ell}_\bp \cdot {\bs \Omega}_{\bp} (\bm A \cdot e {\bs B}).
\eeqn

In Eq.~\eq{eom}, we also define the effective electric field
\beqn
e {\widetilde {\bs E}} = - \frac{\partial H}{\partial \bx} \equiv e {\bs E} -  \frac{\partial \epsilon_\bp}{\partial \bx},
\eeqn
and the effective group velocity,
\beqn
{\tilde \bv}_\bp = \frac{\partial H}{\partial \bp} \equiv \bv_\bp + A_0 \bl.
\eeqn
where ${\bs E} = - \partial  A_0/\partial \bx$ and $\bv_\bp = \partial \epsilon_\bp / \partial \bp$ are the ordinary electric field and the group velocity, respectively. We assume that the system resides in the local thermodynamic equilibrium.

\section{Transport}
\label{sec:transport}

\subsection{Particle density and magnetic helicity}

The particle number density,
\beqn
n & = & \int \frac{d^3 \bp}{(2 \pi)^3}  \sqrt{w} n_\bp  = \int \frac{d^3 \bp}{(2 \pi)^3}  
\label{eq:n:general} \\
  &   & \cdot \Big(1 +  {\bs \ell}_\bp \cdot \bm A +  e {\bs B} \cdot {\bs \Omega}_{\bp} \nonumber
  + {\bs \ell}_\bp \cdot {\bs \Omega}_{\bp} (\bm A \cdot e {\bs B}) \Big) n_\bp,
\eeqn
can be rewritten, in a homogeneous system, as follows:
\beqn
n = \frac{1}{2 \pi^2}  \int_0^{\infty} p^2 n_\bp dp 
     + s \frac{\bm A \cdot \bm B}{2 \pi^2} \int_0^{\infty} e(p) \frac{d e(p)}{dp} n_\bp dp. \qquad
\label{eq:n}
\eeqn
To derive the above relation, we used the explicit expressions of the Berry curvature~\eq{eq:A:Omega} and the conformal vector~\eq{eq:l}, and assumed the isotropy of the occupation number, $n_\bp \equiv n_p$.

Let us consider a zero-temperature case with the sharp boundary of the Fermi fluid at $p = p_F = \mu$ (in this article, we set the Fermi velocity $v_F = 1$ for simplicity). Then the density~\eq{eq:n} may be computed explicitly:
\beqn
n = \frac{\mu^3}{6 \pi^2} + s \frac{\bm A \cdot \bm B}{4 \pi^2} \bigl[e^2(\mu) - e^2(0)\bigr].
\label{eq:n:dens}
\eeqn
The first term is the conventional thermodynamic expression while the second term gives us a new contribution proportional to the local density of magnetic helicity~\eq{eq:AB}. The second term appears due to the conformal anomaly which determines how the electric coupling runs with the energy scale.

Integrating the particle density~\eq{eq:n:dens} over the volume, we get the total particle number at $T=0$:
\beqn
N & = & \frac{\mu^3}{6 \pi^2} V + \frac{s}{4\pi^2} \bigl[e^2(\mu) - e^2(0)\bigr] H \nonumber \\
& = & \frac{\mu^3}{6 \pi^2} V + \frac{s}{\pi} \beta_\alpha H + \dots,
\label{eq:N:total}
\eeqn
where $V$ is the volume of the system and $H$ is the total magnetic helicity~\eq{eq:helicity}. The second line is expressed in the infrared regime at low $\mu$, where $\beta_\alpha$ is the beta function~\eq{eq:beta:alpha} taken at $\mu = 0$. The ellipsis in Eq.~\eq{eq:N:total} denote the higher order terms in the series over~$\mu$. 

Equation~\eq{eq:N:total} implies that a variation of the magnetic helicity $H$ modifies the number of particles $N$ in the ensemble. In the leading order, the number of particles pumped into the ensemble depends linearly on the helicity $H$ with the slope controlled by the beta function. 

\subsection{Axial anomaly and particle current}

Multiplying Eq.~(\ref{eom}) by the factor $\sqrt{w}$ and performing the integral over the momentum $\bm p$, we arrive to the following relation which expresses the nonconservation of the particle number due to the presence of the axial anomaly:
\begin{eqnarray}
\partial_\mu j^\mu & \equiv &
{\dot n} + {\bs \nabla} {\bs j} = - \int \frac{d^3 \bp}{(2 \pi)^3}  e^2(p) ({\bs E} \cdot {\bs B})
\left( {\bs \Omega}_{\bs p} \cdot \frac{\partial n_{\bs p}}{\partial {\bs p}} \right)
\nonumber \\
& = & \frac{s}{2\pi^2} (\bm E \cdot \bm B) \int_0^\infty d p \, e^2(p) \delta(\mu - p)  \nonumber \\
& = & s \frac{{\bs E} \cdot {\bs B}}{2 \pi^2} e^2(\mu).
\label{continuity}
\end{eqnarray}
This expression matches the one in Ref.~\cite{Son:2012zy} with the single modification that we have now taken into account the dependence of the electric coupling on the momentum scale, $e = e(p)$. To pass to the second line of Eq.~\eq{continuity}, we used the sharpness of the Fermi surface at zero temperature: $n_{\bs p} = \theta(\mu - p)$. Therefore, $\partial_{\bs p} n_{\bs p} = - ({\bs p}/p) \, \delta(\mu - p)$ so that $({\bs \Omega}_{\bs p} \cdot \partial_{\bs p} n_{\bs p}) = - s \delta(\mu - p)/p^2$. 

For the scalar particle with the spin $s=0$, the chiral anomaly~\eq{continuity} is evidently absent. For the spin-half fermions, the axial anomaly affects the right-handed and left-handed particles in the opposite manner (with, respectively, $s = +1/2$ and $s = - 1/2$ in our notations) so that the vector charge, given by the sum of these numbers, is conserved. The axial charge, represented via the difference of these numbers, is subjected to the conformal anomaly. Both these statements may be expressed as follows:
\beqn
j^\mu_V = j^\mu_R + j^\mu_L, \qquad j^\mu_A = j^\mu_R - j^\mu_L,
\label{eq:j:VA}
\eeqn
with $\partial_\mu j^\mu_V = 0$ and 
\beqn
\partial_\mu j^\mu_A = - \frac{e \bm E \cdot e \bm B}{2 \pi^2},
\eeqn
where $e \equiv e(\mu)$ is the electric coupling evaluated at the Fermi surface.

The total particle current in Eq.~\eq{continuity} has the following explicit form:
\beqn
{\bm j} & = & - \int \frac{d^3 \bp}{(2 \pi)^3} \Big[  \epsilon_\bp  \frac{\partial n_\bp}{\partial \bm p} - (e  \bm E \times {\bs \Omega}_{\bp}) \ n_\bp \nonumber \\
& & +  \epsilon_\bp   \Big({\bs \Omega}_{\bp} \cdot \frac{\partial n_\bp}{\partial \bm p}\Big) e {\bs B} 
+  \epsilon_\bp  \Big({\bs \Omega}_{\bp} \times \frac{\partial n_\bp}{\partial \bm x}\Big) \nonumber\\
& & - (e \bm E \times \bm A + A_0 e {\bs B}) ({\bs \Omega}_{\bp} \cdot {\bs \ell}_\bp) n_\bp 
\label{jp}
\\
& & +  \epsilon_\bp \bm A \times \Big(\frac{\partial n_\bp}{\partial \bm p} \times {\bs \ell}_\bp \Big) - A_0 {\bs \ell}_\bp n_\bp \nonumber \\
& & + \epsilon_\bp ({\bs \Omega}_{\bp} \cdot {\bs \ell}_\bp) \Big(\bm A \times \frac{\partial n_\bp}{\partial \bm x}\Big)
\Big].
\nonumber
\eeqn
The first term in the right-hand-side of this expression corresponds to the ordinary particle current. The second term gives the anomalous Hall current while the third term determines the standard CME~\cite{SonY:2012}. These terms follow by the fourth contribution which takes into account the spatial inhomogeneities of the system.

The last three lines in Eq.~(\ref{jp}) correspond to the effects of the conformal anomaly. Among those, the very last line has been obtained by integration by parts, taking into account the fact that the following integrals around the Fermi surface vanish:
\beqn
\int \frac{d^3 \bp}{(2 \pi)^3} \partial_\bp(e \bm E n_\bp) & = & 0, 
\label{eq:zero:one} \\
\int \frac{d^3 \bp}{(2 \pi)^3} \partial_\bp \Bigl(e \bm B \cdot\frac{\partial n_\bp}{\partial \bm x}\epsilon_\bp {\bs \Omega}_{\bp}\Bigr) & =& 0. \label{eq:zero:two}
\eeqn
The terms~\eq{eq:zero:one} and \eq{eq:zero:two} come from the first and last terms in the curly parenthesis in the left-hand side of Eq. (\ref{eom}), respectively.

In line the conclusions of Refs.~\cite{SonY:2012,Son:2012zy}, the definitions of the particle density~\eq{eq:n:general} and the particle current~\eq{jp} get modified to include the effects of the Berry curvature. In our case, we also arrive to new terms generated due to the conformal anomaly in both expressions.

Alternatively, the particle current can also be defined in a straightforward manner:
\beqn
 \bm{j}=\int  \frac{d^3 \bp}{(2 \pi)^3} \sqrt{w} \dot{\bm x} n_\bp .
 \label{jbp}
\eeqn
In the absence of the conformal anomaly, $\bl = 0$, the difference between the definitions~\eq{jp} and \eq{jbp} is given by the magnetization term, $\bm j \to \bm j + \bm \nabla \times \bm M$, where
\beqn
 \bm{M} = \int  \frac{d^3 \bp}{(2 \pi)^3} \epsilon_\bp {\bs \Omega}_{\bp} n_\bp,
 \nonumber
\eeqn
is the total magnetization~\cite{Son:2012zy,chen2014:ssyyLI}. These relations are supported by the identity $\bm \nabla \times \bm ({\bs \Omega}_{\bp} \epsilon_\bp n_\bp)=0$ which fixes the ambiguity in the definitions of the current~\cite{Son:2012zy}. 

In our conformally anomalous system, one can also demonstrate that the definitions (\ref{jbp}) and (\ref{jp}) are equivalent up to the magnetization term, which now includes the new contribution generated by the conformal anomaly: $\bm j \to \bm j + \bm \nabla \times [\bm M + \bm A ({\bs \Omega}_{\bp} \cdot {\bs \ell}_\bp) \epsilon_\bp n_\bp ]$.

\subsection{Particle current and helicity current}

Using Eq.~(\ref{eq:EOM:2}), we rewrite the particle current,
\beqn
  {\bm j} & = & \int \frac{d^3 \bp}{(2 \pi)^3} \sqrt{w} \dot{\bm x} n_\bp  
  \label{je} \\
  & = &  \int \frac{d^3 \bp}{(2 \pi)^3} \Bigr( {\tilde \bv}_\bp +  {\bs A} \times [ {\tilde \bv}_\bp \times \bl ] +  e {\widetilde {\bs E}} \times {\bs \Omega}_{\bp}  \nonumber \\
  & & + e {\bs B} ({\bs \Omega}_{\bp} \cdot {\tilde \bv}_\bp) + e ({\bs \Omega}_{\bp} \cdot {\bs \ell}_\bp) {\widetilde {\bs E}} \times \bm A \Bigr) e n_{\bp},\nonumber
\eeqn
which may now be represented as a sum of two terms:\footnote{We use the subscripts ``axial'' and ``conf'' to highlight the particular quantum anomaly responsible for each of these contributions.}
\beqn
{\bs j} & = & {\bs j}_{\mathrm{axial}} 
+ {\bs j}_{\mathrm{conf}}.
\label{eq:jinitial}
\eeqn
The first term in the right-hand side of Eq.~\eq{eq:jinitial} is the single-chirality particle current coming from the axial anomaly:
\beqn
{\bs j}_{\mathrm{axial}} = s {\bm B} \int \frac{d^3 \bp}{(2 \pi)^3} \frac{e}{p^2} n_{\bs p}.
\label{eq:j:axial}
\eeqn
This term gives rise~\cite{CME:SY} to the chiral magnetic effect (CME)~\cite{Kharzeev:2008fw} and the charge separation effect (CSE)~\cite{Metlitski:2005z} that generate the vector and axial currents, respectively:
\beqn
{\bs j}_V = \frac{\mu_A}{2\pi^2} {e \bs B}, 
\qquad 
{\bs j}_A = \frac{\mu_V}{2\pi^2} {e \bs B}.
\label{eq:CME}
\eeqn
Here 
\beqn
\mu_V = \frac{\mu_R + \mu_L}{2}, \qquad \mu_A = \frac{\mu_R - \mu_L}{2},
\label{eq:mu:VA}
\eeqn
are the corresponding chemical potentials.

The second term in Eq.~\eq{eq:jinitial} is new:
\beqn
{\bs j}_{\mathrm{conf}} = s (\bm E \times \bm A+A_0 \bm B) \int \frac{d^3 \bp}{(2 \pi)^3} \frac{e}{p^2} \frac{de}{dp} n_{\bs p}.
\label{eq:j:conf}
\eeqn
The particle current~\eq{eq:j:conf} is generated along the (magnetic) helicity current of the electromagnetic background:
\beqn
{\bs K} = {\bm E} \times {\bm A} + A_0 {\bm B},
\label{eq:K:vector}
\eeqn
The new current~\eq{eq:j:conf} is activated by the conformal anomaly associated with the running of the electric coupling~\eq{eq:beta:g} because if the conformal anomaly is absent, $d e/d p = 0$, then the current~\eq{eq:j:conf} vanishes: ${\bs j}_{\mathrm{conf}} = 0$.

We combine the conformal contribution of the particle density~\eq{eq:n} and the particle current \eq{eq:j:conf} into the 4-vector current of right-handed ($s=+1/2$) and left-handed ($s=-1/2$) particles:
\beqn
j^\mu_{\mathrm{conf}} = K^\mu \frac{s}{2 \pi^2} \int\limits_0^\infty e \frac{d e}{d p} n_p dp,
\label{eq:j:h}
\eeqn
where the Lorentz-covariant expression 
\beqn
K^\mu = \frac{1}{2} \epsilon^{\mu\nu\alpha\beta} A_\nu F_{\alpha\beta},
\label{eq:K}
\eeqn
is known as the Chern-Simons current. It includes the density~\eq{eq:AB} and the 3-current~\eq{eq:K:vector} of the magnetic helicity.

The beta function~\eq{eq:beta:alpha}, integrated along a radius of the Fermi sphere, gives us the strength of the current induced by the conformal anomaly~\eq{eq:j:h} weighted by the occupation number $n_p$. At zero temperature, the Fermi surface is sharp and the integral in Eq.~\eq{eq:j:h} may be done exactly: 
\beqn
j^\mu_{\mathrm{conf}} = \frac{s K^\mu}{(2 \pi)^2} [e^2(\mu) - e^2(0)].
\label{eq:j:conf:T0}
\eeqn
The proportionality factor, which has a sense of the conductivity, takes into account the difference in the values of the electric coupling in the very center of the Fermi volume and at its surface. The difference appears due to the running of the electric coupling with the energy scale.

The zeroth component of this expression coincides with the anomalous part of the density~\eq{eq:n:dens}. In the infrared limit, $\mu \to 0$, the particle current~\eq{eq:j:h} takes a Lorentz-covariant form: 
\beqn
j^\mu_{\mathrm{conf}} = \frac{s \beta_\alpha}{\pi} K^\mu, 
\label{eq:j:mu:concise}
\eeqn
where the beta function~$\beta_\alpha$, Eq.~\eq{eq:beta:alpha}, is evaluated at the tip of the Dirac cone, $\beta_\alpha = \beta_\alpha(\mu=0)$. 

The helicity current~\eq{eq:K} has a gauge-invariant counterpart of the zilch current~\cite{Lipkin:1964} which is a conserved quantity in the Maxwell electrodynamics. Both helicity~\cite{Avkhadiev:2017fxj,Yamamoto:2017uul,Zyuzin} and zilch~\cite{Chernodub:2018era,Copetti:2018mxw,Huang:2020kik,Prokhorov:2020npf} currents can be generated in a rotating hot photon gas  along the axis of rotation. The emergence of these photonic currents is a consequence of the mixed gravitational anomaly for photons which states that the photon helicity is not conserved in a gravitational background~\cite{Dolgov:1988qx}.

The full particle current~\eq{je} is not conserved due to the axial anomaly~\eq{continuity}. The part of the current, corresponding to the contribution of the conformal anomaly, is not conserved as well:
\beqn
\partial_\mu j^\mu_{\mathrm{conf}}(x) & = & \frac{s Q(x)}{2 \pi^2} \int e \frac{de}{dp} n_{\bs p} dp \nonumber \\
& & + \frac{s K^0}{2 \pi^2} \int e \frac{de}{dp} {\dot n}_{\bs p} dp \nonumber \\
& & + \frac{s {\bs K}}{2 \pi^2} \int e \frac{de}{dp} \frac{\partial n_{\bs p}}{\partial {\bs x}} dp.
\label{eq:continuity:h}
\eeqn
The first term in this expression shares similarity with the axial anomaly~\eq{continuity} linked to the non-conservation of the photonic helicity number:
 \beqn
Q = \partial_\mu K^\mu \equiv \frac{1}{2} F_{\mu\nu} {\widetilde F}^{\mu\nu} \equiv - 2 {\bs E}\cdot {\bs B},
\label{eq:Q}
\eeqn
where ${\widetilde F}^{\mu\nu} = \frac{1}{2}\epsilon^{\mu\nu\alpha\beta} F_{\alpha\beta}$ with $\epsilon^{0123} = + 1$. The quantity~\eq{eq:Q} is proportional to the topological charge density of the background electromagnetic field. The evolution of the occupation number ${\dot n}_{\bs p}$ in the second line of Eq.~\eq{eq:continuity:h} can be read off from the equation of motion~\eq{eom}. In a homogeneous system, the last two terms in Eq.~\eq{eq:continuity:h} vanish, and the non-conservation of the conformally-generated particle current takes the following form:
\beqn
\partial_\mu j^\mu_{\mathrm{conf}}(x) & = & \frac{s Q(x)}{4 \pi^2} \bigl[ e^2(\mu) - e^2(0) \bigr] \nonumber \\
& \equiv & - s \frac{{\bs E} \cdot {\bs B}}{2 \pi^2} \bigl[ e^2(\mu) - e^2(0) \bigr],
\label{eq:d:j:conf}
\eeqn
which can also be obtained from Eqs.~\eq{eq:j:conf:T0} and \eq{eq:Q}. Notice the similarity between Eq.~\eq{eq:d:j:conf} and the continuity equation for the full particle current~\eq{continuity}.

\subsection{Vector and axial currents due to conformal anomaly}

The contribution of the conformal anomaly to the vector and axial currents is given by the linear combinations~\eq{eq:j:VA} of the right-handed ($s = + 1/2$) and left-handed ($s = - 1/2$) particle currents~\eq{eq:j:conf:T0}. The corresponding chemical potentials, $\mu_R$ and $\mu_L$, affect the conformal currents via the running of the electric coupling $e(\mu_{R,L})$. In the presence of the chiral imbalance, $\mu_R \neq \mu_L$, the Fermi surfaces do not coincide with each other, the electric couplings do not match (Fig.~\ref{fig:illustration}), and the corresponding currents acquire an additional dependence on the axial chemical potential~$\mu_A$, Eq.~\eq{eq:mu:VA}. This mismatch in the Fermi surfaces gives rise to additional contributions to the vector and axial currents as we discuss below.

To keep our expressions concise, we will only show the leading term in the series over the powers of the chemical potentials. If the coupling constant changes substantially -- as we move from the tip of the Dirac cone ($p=0$) to the Fermi surface ($p_F = \mu$) -- then the beta function should be replaced by 
\beqn
\beta_\alpha \to \frac{1}{4 \pi} [e^2(\mu) - e^2(0)].
\label{eq:substitution}
\eeqn

In the case of zero axial imbalance, $\mu_A = 0$, and $\mu_R = \mu_L = 
\mu_V$, the conformal anomaly~\eq{eq:j:conf:T0} leads to the following currents:
\beqn
\left(j^\mu_V\right)_{\mathrm{conf}} = 0, \quad \left(j^\mu_A\right)_{\mathrm{conf}} = \frac{\beta_\alpha}{\pi} K^\mu.
\qquad
\label{eq:j:muA0}
\eeqn
The vector current vanishes, while the axial current is expectedly proportional to the helicity flow of the electromagnetic background field. The same result~\eq{eq:j:muA0} is true for a different physical environment, where the vector chemical potential is zero, $\mu_V = 0$ while $\mu_R = - \mu_L = \mu_A$. The reason for this coincidence is that the running of the coupling is insensitive to the sign of the chemical potential due to the charge-conjugation symmetry that puts the equivalence between particles and holes (anti-particles): $e(\mu) = e(- \mu)$. Notice also that the electric coupling $e$ enters the anomalous particle current~\eq{eq:j:h} as $e^2$ which is invariant under the sign flip, $e \to - e$.

The independence of the axial current~\eq{eq:j:muA0} on the chemical potentials implies that this current may emerge at smallest deviations from the neutrality point at the tip of the Dirac cone where all chemical potentials are close to zero. 

The proportionality of the axial particle current $j^\mu_A$ to the helicity current $K^\mu$ is a natural outcome from the point of view the discrete $C$-, $P$-, and $T$- symmetries that are shared by the both currents. The dependence~\eq{eq:j:muA0} has a suggestive form if we compare it with the axial anomaly equation~\eq{continuity} which may be written in terms of the conservation law:
\beqn
\partial_\mu \Bigl(j^\mu_A - \frac{e^2}{4 \pi^2} K^\mu\Bigr) = 0.
\label{eq:d:jA:K}
\eeqn
While this equation states that the helicity of light is transferred to the chiral fermions and vice-versa, it does not imply the local equivalence or proportionality of these both currents. Our result~\eq{eq:j:muA0}, together with Eq.~\eq{eq:substitution}, indicates that the axial current does pick a part of the helical current if the electric coupling runs with the energy scale. 

Another physically interesting case corresponds to a finite-density system ($\mu_V \neq 0$) characterized by a small chiral imbalance, $|\mu_A| \ll |\mu_V|$. The leading contributions to the vector and axial current are, respectively, as follows:
\beqn
\left(j^\mu_V\right)_{\mathrm{conf}} = \frac{\beta_\alpha}{\pi} \frac{\mu_A}{\mu_V} K^\mu, 
\qquad 
\left(j^\mu_A\right)_{\mathrm{conf}}  = \frac{\beta_\alpha}{\pi} K^\mu.
\label{eq:j:muV:muA}
\eeqn
Interestingly, the vector current also gets a contribution proportional to the electromagnetic helicity flow with the strength controlled by the beta function $\beta_\alpha$ and the axial chemical potential $\mu_A$. Due to the invariance of Eq.~\eq{eq:j:muV:muA} under time reversal, the axial and vector currents are dissipationless quantities. The conformally induced currents have the same symmetries under parity and time inversions as the currents generated by the dissipationless CME and CSE effects~\eq{eq:CME}. 

The discussed transport effects of the conformal anomaly in the matter are different from the scale electromagnetic phenomenon in the vacuum~\cite{Chernodub:2016lbo}. The vacuum effects are realized in the gravitational background while the effects discussed in this paper occur in a flat spacetime. In fact, the scale magnetic effects cannot be seen in our approach due to two reasons. First of all, we are working in the $O(\hbar)$ order of the kinetic theory, while the SME appears at the $O(\hbar^2)$ order. Secondly, the SME is the vacuum effect, while the chiral kinetic theory, employed in this article, deals with phenomena generated in the presence of matter.

The physical realization of the electromagnetic background with a nonvanishing helical current is provided by the circularly polarized light which has already been discussed in the context of Dirac and Weyl semimetals in Refs.~\cite{ref:circ:1,ref:circ:2}.

\section{Conclusions}

The axial anomaly is known to lead to the dissipationless transport phenomena known as the chiral magnetic and chiral separation effects that generate, respectively, vector and axial currents of fermions along the axis of the background of the magnetic field~\eq{eq:CME}. These effects occur in the presence of matter, either at a chiral imbalance or at finite density, respectively. 

In this article, we demonstrate that the conformal anomaly, associated with the running of the electric coupling~\eq{eq:beta:alpha} in the presence of fermionic matter, can also lead to the appropriate dissipationless transport, which is different from the currents generated by the axial anomaly. The transport effects of the conformal anomaly are activated in the background electromagnetic fields, which possess a nonvanishing flux of magnetic helicity~\eq{eq:K}. 

Working in the collisionless regime of the chiral kinetic theory, we show that the conformal anomaly generates the axial (chiral) current along the helicity current of the electromagnetic background field~\eq{eq:j:muA0}. The anomalous current's strength depends on the difference in the renormalized electric couplings in the very center of the Fermi volume and at the Fermi surface~\eq{eq:j:conf:T0}. This difference is determined by how the electric coupling runs with the renormalization group and is related to its beta function and, thus, to the conformal anomaly. In the vicinity of the neutrality point, the conductivity depends on the beta function only. In a finite-density environment, the conformal anomaly also induces the vector current of particles along the direction of the flux of the magnetic-helicity~\eq{eq:j:muV:muA}.

The discussed conformal transport effects emerge in matter provided the combinations $e B$ and $e E$ of the electromagnetic background $(E, B)$ and the electric coupling $e$ are not renormalization-group invariants. While these combinations do not run with the interaction scale in the fundamental QED, we argued that this is not a generic case, especially for models that describe the infrared properties of solid-state systems. For simplicity of presentation, we assumed that the electromagnetic background is not renormalized by quantum corrections.

Similarly to the axial anomaly, the conformal anomaly reveals itself as a monopole in the momentum space~\eq{eq:l}. Contrary to the axial counterparts, the discussed conformal transport effects do not have a topological origin and, therefore, they are not one-loop exact.

\begin{acknowledgments}
M.C. is partially supported by Grant No. 0657-2020-0015 of the Ministry of Science and Higher Education of Russia. E.K. is grateful to the Embassy of France in Turkey for the French Embassy Research Fellowship which supported her visit to the Institut Denis Poisson where this work was initiated. 
\end{acknowledgments}

\end{document}